\documentclass[aps,twocolumn,superscriptaddress,floatfix,showpacs,10pt,letterpaper,nofootinbib]{revtex4-1}

\usepackage{graphicx}
\usepackage{graphics} 
\usepackage{ifpdf}
\ifpdf
\usepackage{epstopdf}
\usepackage{hyperref}
\else
\usepackage[hypertex]{hyperref}
\fi
\usepackage{xypic}

\input xypic
\usepackage[all,tips]{xy}

\allowdisplaybreaks[1]

\newlength{\fighskip} \fighskip=2pt
\newlength{\figvskip} \figvskip=3pt

\newcommand\void[1]       {}

\usepackage{amsthm}
\usepackage{euscript}
\usepackage{epstopdf} 

\newtheoremstyle{wenthm}
  {3pt}
  {3pt}
  {\slshape}
  {}
  {\bfseries}
  {:}
  {.5em}
  {}

\theoremstyle{wenthm}

\newtheorem{prop}{Result}

\theoremstyle{definition}

\newtheorem{defn}{Definition}

\newtheorem{app}{Application}

\begin{document}

\begin{titlepage}

\title{A relation between chiral central charge and ground state degeneracy\\
in  2+1-dimensional topological orders}

\author{Liang Kong}
\affiliation{
Guangdong Provincial Key Laboratory of Quantum Science and Engineering,
and Shenzhen Institute for Quantum Science and Engineering,
Southern University of Science and Technology, Shenzhen 518055, China
}

\author{Xiao-Gang Wen} 
\affiliation{Department of Physics, Massachusetts Institute of Technology, Cambridge, Massachusetts 02139, USA}

\begin{abstract} 

A bosonic topological order on $d$-dimensional closed space $\Sigma^d$ may have
degenerate ground states.  The space $\Sigma^d$ with different shapes
(different metrics) form a moduli space ${\cal M}_{\Sigma^d}$.  Thus the
degenerate ground states on every point in the moduli space ${\cal
M}_{\Sigma^d}$ form a complex vector bundle over ${\cal M}_{\Sigma^d}$.  It was
suggested that the collection of such vector bundles for $d$-dimensional closed
spaces of all topologies completely characterizes the topological order.  Using
such a point of view, we propose a direct relation between two seemingly
unrelated properties of 2+1-dimensional topological orders: (1) the chiral
central charge $c$ that describes the many-body density of states for edge
excitations (or more precisely the thermal Hall conductance of the edge), (2)
the ground state degeneracy $D_g$ on closed genus-$g$ surface.  We show that $c
D_g/2 \in \mathbb{Z},\ g\geq 3$ for bosonic topological orders.  We explicitly
checked the validity of this relation for over 140 simple topological orders.
For fermionic topological orders, let $D_{g,\sigma}^{e}$ ($D_{g,\sigma}^{o}$)
be the degeneracies with even (odd) number of fermions on genus-$g$ surface
with spin structure $\sigma$. Then we have $2c D_{g,\sigma}^{e} \in \mathbb{Z}$
and $2c D_{g,\sigma}^{o} \in \mathbb{Z}$ for $g\geq 3$.

\end{abstract}


\maketitle

\end{titlepage}

\setcounter{tocdepth}{1} 
{\small \tableofcontents }

\section{Introduction}
\label{intro}

In 1989, through a theoretical study of chiral spin
liquid,\cite{KL8795,WWZ8913} we realized that there exists a new kind of order
-- topological order\cite{Wtop,WNtop,Wrig} -- beyond Landau symmetry breaking
theory.  Topological order cannot be characterized by the local order
parameters associated with the symmetry breaking.  However, topological order
can be characterized/defined by the following macroscopic properties: (a) the
topology-dependent ground state degeneracy\cite{Wtop,WNtop} and (b) the
non-Abelian geometric phases of the degenerate ground states as we deform the
metrics of the space.\cite{Wrig,KW9327} Both of the above macroscopic
properties are \emph{robust against any local perturbations} that can break any
symmetries.\cite{WNtop} This is just like superfluid order which is
characterized/defined by zero-viscosity and quantized vorticity that are robust
against any local perturbations that preserve the $U(1)$ symmetry.  

For topological orders in 2+1-dimensional (2+1D) spacetime, the non-Abelian
geometric phases of the degenerate ground states encode the chiral central
charge $c$ of the topological order via the gravitational Chern-Simons term in
the effective action.\cite{AG14013703,GF14106812}  (The edge of a topological
order contains right-movers with central charge $c_R$, and  left-movers with
central charge $c_L$.  The chiral central charge
$c=c_R-c_L$.\cite{W9038,KF9732}) In this paper, we propose a direct relation
between the ground state degeneracy $D_g$ on a genus-$g$ space and the chiral
central charge $c$ of the edge states, for 2+1D topological orders in bosonic
systems:
\begin{align}
 \frac12 c D_g \in \Z \ \ \text{ for } g\geq 3.
\end{align}
For 2+1D topological orders in fermionic systems, we propose
\begin{align}
 2 c D_g \in \Z \ \ \text{ for } g\geq 3.
\end{align}
This result can be derived from the characterization of a topological order in
terms of a complex vector bundle on the moduli space $\cM_{\Si^d}$ of a closed
space $\Si^d$,  where the fiber is the degenerate ground states on  $\Si^d$.
We also discuss how to use the partition function on arbitrary closed spacetime
$M^D$ and the resulting complex function on $\cM_{M^D}$, the moduli space of
$M^D$, to  characterize topological orders.

In this paper, we will use $d$ or 1d, 2d, \etc\ to denote the space dimension
and $D$ or 1+1D, 2+1D, \etc\ to denote the spacetime dimension.  We will only
consider anomaly-free topological orders.

\section{Probe and measure the topological orders -- Hamiltonian
approach }

The main issue that we are going to discuss in this paper is how to probe and
measure different topological orders using macroscopic properties.  Here
``probe and measure'' means the methods in experiments and/or numerical
calculations that allow us to distinguish different topological orders.  

\subsection{Complex vector bundle on moduli space of space manifold}

In Hamiltonian approach, an anomaly-free topological order is described by a local
bosonic Hamiltonian acting on a Hilbert space $\cV_\text{tot}$ with a tensor
product decomposition
\begin{align}
 \cV_\text{tot} =\bigotimes_i \cV_i
\end{align}
where $\cV_i$ is the finite-dimensional Hilbert space for site-$i$.  The
Hamiltonian is required to be well defined for arbitrary space $\Si^d$ with
arbitrary triangulation.  
Since anomaly-free  topological orders are gapped, we require the Hamiltonian on a
closed space $\Si^d$ to be gapped, whose degenerate ground states form a finite
dimensional vector space $\cV_\text{grnd}$, which is a subspace of the total
Hilbert space $\cV_\text{tot}$ of the boson system.  Let $\cM_{\Si^d}$ be the
moduli space for closed space $\Si^d$ with different metrics and $\cM$ the
disjoint union of these moduli spaces.  We see that there is a ground-state
vector space $\cV_\text{grnd}$ for every point in $\cM_{\Si^d}$. Therefore, for
each $\Si^d$, an anomaly-free topological order gives rise to a complex vector bundle $\cE_{\Si^d}$ with fiber $\cV_\text{grnd}$ and base space $\cM_{\Si^d}$.
It was propose in \Ref{Wrig} that
\begin{prop} 
the complex vector bundle
\begin{align}
 pt \to \cV_\text{grnd} \to \cE_{\Si^d} \to \cM_{\Si^d} \to pt
\end{align}
of degenerate ground states on $\cM_{\Si^d}$ (for all $\Si^d$'s) fully
characterize an anomaly-free topological order in Hamiltonian
formalism.
\end{prop}

To understand the vector bundle $\cE_{\Si^d}$, let us use
$G_\text{homeo}(\Si^d)$ to denote the orientation preserving homeomorphism
group of the space $\Si^d$.  Note that $G_\text{homeo}(\Si^d)$ only depends on
the topology of $\Si^d$  and is the same for every point $\Si^d \in
\cM_{\Si^d}$.  Let us use $G^0_\text{homeo}(\Si^d)$ to denote the subgroup of
$G_\text{homeo}(\Si^d)$ which is the connected component of
$G_\text{homeo}(\Si^d)$ that contain identity.  The mapping class group is
formed by the discrete components of the homeomorphism group:
\begin{defn} \textbf{mapping class group}\\
$\text{MCG}(\Si^d)\equiv  G_\text{homeo}(\Si^d)/G^0_\text{homeo}(\Si^d)=\pi_0[G_\text{homeo}(\Si^d)]$.
\end{defn} \noindent
We note that every homeomorphism $g: \Si^d\to \Si^d$ in $\text{MCG}(\Si^d)$
defines a mapping torus $\Si^d \gext_g S^1$ that describes how $\Si^d$ deform
around a loop $S^1$, and corresponds to an element in $\pi_1(\cM_{\Si^d})$.
Thus $\pi_1(\cM_{\Si^d})=\text{MCG}(\Si^{d})$.  

Along a loop $g$ in $\pi_1(\cM_{\Si^{d}})$, the fiber bundle gives us a
monodromy $U(g)$ which is a unitary matrix acting on the ground state vector
space $\cV_\text{grnd}$.  We may view $g$ as an element in the group
$\text{MCG}(\Si^{d})$.  So $U(g)$ gives a projective representation of
$\text{MCG}(\Si^{d})$.\cite{W8951}

To understand why we only get a projective representation, we note that the
topological robustness of the ground state degeneracy implies that the unitary
matrix $U(\mathrm{id})$ for a contractible loop $g=\mathrm{id}$ in the moduli
space $\cM_{\Si^d}$ must be a pure over-all phase (which can be path
dependent),  so that $U(\mathrm{id})$ cannot distinguish (or split) the
degenerate ground states.  This is because the periodic time evolution along a
contractable loop over and over again can be simulated by a local Hamiltonian.
If $U(\mathrm{id})$ can distinguish the degenerate ground states, 
then there is a local Hamiltonian that can distinguish and split the degenerate
ground states. This contradicts with the fact that the topological degeneracy of
ground states cannot be lifted by any local Hamiltonian.  Similarly, $U(g)$ may
also depend on paths, but the path-dependent part must be an over-all phase.
This leads to the projective representation of $\text{MCG}(\Si^{d})$.  We also
like to mention that the trace of $U(g)$ is the volume-independent partition
function (see Section \ref{PMtopo} for details) on the corresponding mapping
torus: 
\begin{align}
\label{UgZ}
\Tr \, U(g) =Z^\text{top}(\Si^{d} \gext_g S^1) .  
\end{align}
As a result, we obtain 
\begin{align}
|Z^\text{top}(\Si^{d} \times S^1)| =\text{ground state degeneracy on }\Si^d .  
\end{align}

For spaces with different topologies, we get different projective
representations.  Those finite dimensional projective representations are the
non-Abelian geometric phases of the degenerate ground states introduced in
\Ref{Wrig,KW9327}.  Certainly, the non-Abelian geometric phases contain more
information than the projective representations. They contain all the
information about the vector bundle $\cE_{\Si^d}$ on $\cM_{\Si^d}$. Therefore,
we believe that such geometric phases for closed $d$-dimensional spaces $\Si^d$
of all topologies fully characterize the topological order.

We like to remark that for a generic system, its vector bundle $\cE_{\Si^d}$
is generaly not flat. The curvature of the vector bundle can change as we
deform the Hamiltonian locally.  However, for some topological orders, its
vector bundle $\cE_{\Si^d}$ cannot be made flat no matter how we fine tune the
Hamiltonian.  In this case, the  vector bundle $\cE_{\Si^d}$ is topologically
different from a flat bundle.  In this case, the volume-independent partition
function on mapping torus $Z^\text{top}(\Si^{d} \gext_g S^1)$ cannot be
topological (\ie  the volume-independent partition function cannot be a
constant on a connected piece of the moduli space $\cM_{\Si^d}$).  It must
depend on the metrics of the space-time $\Si^d \gext_g S^1$.  

It is very strange since the bosonic system has short range correlation and a
finite energy gap.  In the thermodynamical limit, the space-time becomes flat,
and the bosonic system should not be able to sense the geometry of the
space-time.  The fact that the partition function does depend on the metrics of
the space-time means that the entanglement in the ground state can still sense
the geometry of the space in the flat limit.  We like to link such a geometric
sensitivity to the gapless nature of boundary excitations and entanglement
spectrum:
\begin{prop} 
\label{flatVB}
The ground state vector bundle $\cE_{\Si^d}$ over $\cM_{\Si^d}$ can be deformed
into a flat bundle if and only if the boundary of the corresponding anomaly-free
topological order is gappable.
\end{prop}\noindent

What is the obstruction that prevent the vector bundle to be flat?  First, for
a contractible loop $g=\mathrm{id}$, $U(\mathrm{id})$ is a pure $U(1)$ phase.
So the non-flat part is only contained in the $U(1)$ phase of the complex
vector bundle.  We can examine it by considering the determinant line bundle
$\cE^\text{det}_{\Si^d}$ of the vector bundle $\cE_{\Si^d}$, which is a complex
line bundle over $\cM_{\Si^d}$.  

To connect the determinant line bundle to the partition fuction of the system,
let us consider a contractable loop $S^1$ in $\cM_{\Si^d}$.  We have mentioned
that the partition function for the spacetime $\Si^d \times S^1$ is given by
the monodromy $U(\mathrm{id})=\ee^{\ii \th}$ along the loop (see \eqn{UgZ} and
remember that $U(\mathrm{id})$ is pure phase factor). Therefore the partition
function of the system on spacetime $\Si^d \times S^1$ (see \eqn{gCS3} for an
example) is given by
\begin{align}
 Z^\text{top}(\Si^d \times S^1) &  = D_{\Si^d}\ee^{\ii \th}
\nonumber\\
\text{or }\ [Z^\text{top}(\Si^d \times S^1)]^{D_{\Si^d}} &= D_{\Si^d}^{D_{\Si^d}} \text{Det}
U(\mathrm{id}),
\end{align}
where $D_{\Si^d}=\text{Dim} U(\mathrm{id})$ is the ground state degeneracy on
the closed space $\Si^d$, and $\text{Det} U(\mathrm{id})$ is the monodromy of
the  determinant line bundle around the loop $S^1$.  Now, let us assume that
the loop $S^1$ is the boundary of a 2-dimension submanifold $B \subset
\cM_{\Si^d}$: $S^1 =\prt B$.  We can rewrite that above as
\begin{align}
\label{ZCP}
[Z^\text{top}(\Si^d \times S^1)]^{D_{\Si^d}} &\propto  \text{Det}
U(\mathrm{id})
=\ee^{\ii 2\pi \int_B C}
\nonumber\\
& 
=\ee^{\ii 2\pi D_{\Si^d}\int_{\Si^d\gext B} P}
,
\end{align}
where $C$ is the curvature tensor on the moduli space $\cM_{\Si^d}$ for the
determinant line bundle.  The next expression $\ee^{\ii 2\pi D_{\Si^d} \int_{\Si^d\gext
B} P}$ is motivated by noticing that $Z^\text{top}(\Si^d \times S^1)$ is given
by a gravitational Chern-Simon term $\om$
\begin{align}
 Z^\text{top}(\Si^d \times S^1) \propto
\ee^{\ii  \int_{\Si^d \times S^1} \om}
\end{align}
We can rewrite the above in term of a linear combination of Pontryagin class
on $\Si^d \gext B$
\begin{align}
 Z^\text{top}(\Si^d \times S^1) \propto
\ee^{\ii  \int_{\Si^d \times S^1} \om} =
\ee^{\ii \int_{\Si^d \gext B} P}
\end{align}
since $\prt (\Si^d \gext B) = \Si^d \times S^1$.  Here $\Si^d\gext B$ is a fiber
bundle with the space $\Si^d$ as the fiber and $B$ as the base manifold.  Also
$P= \dd \om$ is a linear combination of Pontryagin class on $\Si^d \gext B$: 
\begin{align}
 P &= \sum_{n_1,n_2,\cdots} \ka_{n_1n_2\cdots} P_{n_1n_2\cdots}
\nonumber\\
P_{n_1n_2\cdots} &= p_{n_1}p_{n_2}\cdots, \ \ \ \
\ka_{n_1n_2\cdots} \in \Q,
\end{align}
and $p_n$ is the $n^\text{th}$ Pontryagin class.  This leads to the expression
\eqn{ZCP}. This is a key assumption in this paper.

Now, let us shrink the loop $S^1$ to a point and $B$ becomes a closed
2-dimensional submanifold in $ \cM_{\Si^d}$.  Then, $\int_B C$ becomes the
Chern number of the line bundle $\cE^\text{det}_{\Si^d}$ on $B$, which always
is an integer. We obtain
\begin{align}
\label{DPon}
 D_{\Si^d} \int_{\Si^d\gext B} P =\text{integer} .
\end{align}
This expression gives us a constraint between the ground state degeneracy
$D_{\Si^d}$ and the gravitational Chern-Simons term in the effective theory.
It is the main result of this paper.  We remark that the above result is
obtained with an assumption that the ground states of topological order can be
put the closed space $\Si^d$  without the need to create some topological
excitations.  Otherwise, the above can still be valid if we set $D_{\Si^d}=0$
when we have to create  topological excitations.  We also like to point out the
first Chern class $C$ (\ie the collection Chern numbers $\int_B C$ for all
closed 2-dimensional subspaces of $ \cM_{\Si^d}$) completely classify the line
boundle $\cE^\text{det}_{\Si^d}$.

Let us consider an example of 2d theory whose gravitational response contains
the gravitational Chern-Simons term:
\begin{align}
\label{gCS3}
 Z^\text{top}(\Si^2\gext S^1) = 
\ee^{\ii \frac{2\pi c}{24} \int_{\Si^2\gext S^1} \om_{3} }
,\ \ \
\dd \om_3 =p_1,
\end{align}
where $c$ is the chiral central charge of the edge states.
For such a theory, \eqn{DPon} becomes
\begin{align}
\frac{c}{24} D_g \int_{\Si^2\gext B^2} p_1 = \text{integer},
\end{align}
for any surface bundle $\Si^2\gext B^2$, where $D_g$ is the ground state
degeneracy on $\Si^2$,  and $g$ is the genus of $\Si^2$.  

Since $\int_{\Si^2\gext B^2} p_1 \neq 0$ for some surface bundle, $\int_{B^2} C
\neq 0$ for some $B$ and the vector bundle $\cE_{\Si^d}$ is not flat if $c\neq
0$.  So the appearance of the gravitational Chern-Simons term implies that the
vector bundle $\cE_{\Si^d}$ is not flat.

It was shown that $\int_{\Si^2\gext B^2} p_1 =0$ mod 12 for any orientable
surface bundles.\cite{CFT1275,GMT0759} If the genus of the fiber $\Si^2$ is
equal or less than 2, then  $\int_{\Si^2\gext B^2} p_1
=0$.\cite{CFT1275,E0612,E9815} If the genus of the fiber $\Si^2$ is equal or
greater than 3, then we can always find a base manifold $B^2$ with a genus
equal or less than 111, such  that there is a surface bundle $\Si^2\gext B^2$
with $\int_{\Si^2\gext B^2} p_1 =\pm 12$.\cite{E9815} Therefore,\footnote{Result
\ref{cDgB} was first obtained in a long unpublished paper \Ref{KW1458}.  This
paper simplifies and extendeds the result in \Ref{KW1458}.}
\begin{prop} 
\label{cDgB}
for a 2d bosonic topological orders, the chiral central charge of the edge
state is quantized as $c D_g/2\in \Z$ for $g\geq 3$, where $D_g$ is the ground
state degeneracy on genus-$g$ space.
\end{prop}\noindent
The above result implies that the chiral central charge $c$ is a rational number.

Let us give some non-trivial checks for Result \ref{cDgB}.
\begin{app} 
For a bosonic quantum Hall state with one branch of edge mode (\ie
$c=1$), the ground state degeneracy $D_g$ must be even for $g\geq 3$.  
\end{app} 
\begin{app} 
For a 2d bosonic topological order, we can use $i,j,k$ to label
the topological excitations.  
The fusion rule is given by $i\otimes j = \oplus_k
N^k_{ij} k$. The ground state degeneracy $D_g$ is then given by\cite{BW0932}
\begin{align}
 D_g&=\Tr \Big(\sum_i N_i N_{\bar i} \Big)^{g-1}
= \sum_i S_{1i}^{-2(g-1)}
\nonumber\\
&
=\big(\sum_i d_i^2\big)^{g-1} \sum_i d_i^{-2(g-1)}
\end{align}
where $\bar i$ is the antiparticle of $i$, the matrix $N_i$ is given by
$(N_i)^k_j=N_{ij}^k$, and $d_i$ is the quantum dimension (the largest
eigenvalue of $N_i$).  Also $S_{ij}$ is the matrix elements of the $S$-matrix
that characterizes the topological order.\cite{Wrig,KW9327} We have $S_{1i} =
\frac{d_i}{\sqrt{\sum_i d_i^2}}$

For filling fraction $\nu=1$ bosonic Pfaffian quantum Hall state, we have
$(d_i) =(1,1,\sqrt 2)$.
We find that $D_1=3$, $D_2=10$, $D_3=36$, $D_4=136$, $D_5= 528$, \etc.
Therefore the chiral central charge must be quantized as $c=0$ mod $1/2$, which
agrees with $c=3/2$.  We also see that $c D_g/2=$ integer is not valid for
$g=2$. 

For the Fibonacci topological order with $(d_i) =(1,\frac{\sqrt5+1}{2})$ and
$c=14/5$, we find that $D_1=2$, $D_2=5$, $D_3=15$, $D_4=50$, $D_5= 175$, \etc.
Indeed, $c D_g/2=$ integer for $g\geq 3$.  We explicitly checked over 140
simple topological orders listed in \Ref{W150605768}, and find that Result
\ref{cDgB} is valid for those bosonic topological orders.

\end{app} 

\begin{app} 
The chiral central charge of 2d invertible anomaly-free topological order is
quantized as $c=0$ mod 2, since $D_g=1$.  A known 2d invertible anomaly-free 
topological order is the $E_8$ state, which has $c=8$.  At the moment, we do
not know if the minimal chiral central charge $c=2$ can be realized by a
2d invertible anomaly-free topological order.  
\end{app} 

If we have a fermionic system, both $\Si^d$ and $\Si^d\gext B^2$ should
be chosen to be spin manifolds.  In this case $\Si^d$ can have a spin
structure, denoted as $\si$, which can be extended to $\Si^d\gext B^2$. The
ground states on $\Si^d$ can carry even or odd numbers of fermions.  We denote
the ground state  degeneracy with even fermions as $D_{\Si^d,\si}^e$ and that
with odd fermions as $D_{\Si^d,\si}^o$.  We note that the even and odd sectors
do not mix due to the conservation of fermion number parity.  Therefore, we
have two vector bundles on the modular space $\cM_{\Si^d}$.

In 2-dimensional space ($d=2$), when $\Si^2\gext B^2$ is spin, we have
$\int_{\Si^2\gext B^2} p_1 =0$ mod 48 for any spin surface
bundles.\cite{CFT1275,E0612} Assuming that $\int_{\Si^2\gext B^2} p_1 =\pm 48$
can be realized for some surface bundle $\Si^2\gext B^2$ if the genus of
$\Si^2$ is greater than $2$, we find that 
\begin{prop} 
\label{cDgF}
For fermionic  topological orders,
the chiral central charge is quantized as 
\begin{align}
2c D_{g,\si}^e\in \Z,\ \ \ 
2c D_{g,\si}^o\in \Z,\ \ \ \ \
 g\geq 3,  
\end{align}
where $D_{g,\si}^e$ ($D_{g,\si}^e$) is the ground state degeneracy on closed
genus-$g$ surface with spin structure $\si$ and even (odd) number of fermions.
\end{prop}\noindent
For fermionic  invertible topological orders, we have
$D_{g,\si}^e+D_{g,\si}^e=1$ and the chiral central charge is quantized as $c=0$
mod 1/2.  The minimal chiral central charge $c=1/2$ for fermionic  invertible
topological orders can be realized by $p+\ii p$ superconductor, which indeed
contain no non-trivial topological excitations.

For a 2d fermionic topological order, the quantum dimensions of excitations
appear pairs of equal values: $d_{2i} = d_{2i+1}$\cite{LW150704673}.  Many 2d
fermionic topological orders are stacking of a fermionic trivial product state
and bosonic topological orders with quantum dimensions $d^B_i$.  In this case,
we either have $D_{g,\si}^e \neq 0,\ D_{g,\si}^o=0$ or $D_{g,\si}^e = 0,\
D_{g,\si}^o \neq 0$.  The total ground state degernacy $D_{g,\si}=D_{g,\si}^e +
D_{g,\si}^o$ is indenpendent of spin structure.  To compute $D_{g,\si}$, we
note that the quantum dimensions for the resulting fermion topological order
are given by $d_{2i}=d_{2i+1}=d^B_i$, and the ground state degeneracies are the
same as the corresponding bosonic topological order: $D_{g,\si}
=D_{g,\si}^e+D_{g,\si}^o =\Big( \sum_i (d_i^B)^2\Big)^{g-1} \sum_i (d_i^B)
^{-2(g-1)} $, and we obtain
\begin{align}
D_{g,\si} &=D_{g,\si}^e+D_{g,\si}^o =\Big(\frac12 \sum_i d_i^2\Big)^{g-1} \frac12 \sum_i d_i^{-2(g-1)} .
\end{align}
Amazingly, when we apply the above formula to more general fermionic
topological orders obtained in \cite{LW150704673}, the above expression always
give us integers which satisfy $2c D_{g,\si}\in \Z$ for $g\geq 3$.  

We like to remark that for fermionic topological orders, $D_{\Si^d,\si}^e$ and
$D_{\Si^d,\si}^o$ may depend on spin structure $\si$ (see
\Ref{PY161209298,WY180105416}).  It is not clear whether
$D_{\Si^d,\si}^e+D_{\Si^d,\si}^o$ depends on spin structure or not. For the
examples examined in \Ref{PY161209298,WY180105416},
$D_{\Si^d,\si}^e+D_{\Si^d,\si}^o$ does not depend on spin structure.

\subsection{No non-trivial bosonic topological order in 1d space}

Next let us consider bosonic 1$d$ topological orders.  Since $\text{MCG}(S^1)$
is trivial, $\cM_{\Si^1}$ is simply connected.  Since the Pontryagin classes
for circle bundle $S^1\gext B$ all vanishes, the determinant bundle of the
vector bundle $\cE^\text{det}_{\Si^1}$ over $\cM_{\Si^1}$ can always be
deformed into a flat one.  Thus the vector bundle $\cE_{\Si^1}$ can be flat.
Such a vector bundle is always trivial since $\cM_{\Si^1}$ is simply connected.
Therefore, all bosonic anomaly-free 1$d$ topological orders are trivial.

It appears that the vector bundle $\cE_{\Si^d}$ on $\cM_{\Si^d}$ is a high
resolution characterization of the anomaly-free topological order.  The
non-trivial anomaly-free topological order should lead to a non-trivial vector
bundle $\cE_{\Si^d}$.  On the other hand, since the structure of the vector
bundle can be so rich, it is very likely that not every allowed  vector bundle
$\cE_{\Si^d}$ on $\cM_{\Si^d}$ can be realized by  anomaly-free topological
orders.

\subsection{How to probe and measure the boundary-gappable topological orders }

For an boundary-gappable topological order, the vector bundle on $\cM_{\Si^d}$ can always
be deformed into a flat one.  In fact, the  boundary-gappable topological orders can be
realized by renormalization-group fixed-point Hamiltonians, which are formed by
commuting projectors\cite{LW0510}, or by renormalization-group fixed-point
Lagrangians which are re-triangulation invariant\cite{TV9265,C161007628} (see
next Section).  The vector bundles on $\cM_{\Si^d}$ obtained from those
fixed-point systems are always flat, and the partition functions on mapping
torus are always topological (which are the state-sum topological
invariants\cite{TV9265,C161007628}).  
For a flat vector bundle, the unitary matrices $U(g)$ (the monodromies for
non-contractible loops) form a representation (instead of a projective
representation) of the mapping class group $\text{MCG}(\Si^{d})$ which fully
characterize the flat bundle:\cite{W9039}
\begin{prop} 
An boundary-gappable topological order is fully characterized by a collection of
representations of the mapping class groups $\text{MCG}(\Si^{d})$ for various
spatial topologies.
\end{prop}\noindent
In particular, the representations of $\text{MCG}(\Si^{d})$ can be computed via
the universal wave function overlap\cite{HW1339,MW1418,HMW1457} or tensor
network calculations.\cite{ZGT1251,TZQ1251,ZMP1233,CV1308}

For 2d boundary-gappable topological orders, the representations of the mapping class
group $\text{MCG}(\Si_1)$ for genus-1 torus are called the modular data, which
already carry a lot of information about the topological
orders\cite{KW9327,RSW0777,W150605768}.  However, recently in
\Ref{MS170802796}, Mignard and Schauenburg found some different topological
orders that have the same modular data and chiral central charge.  Thus,
modular data and chiral central charge are not enough to fully characterize
topological order.  In \Ref{WW190810381}, its was shown that if we include the
representations of the mapping class group  $\text{MCG}(\Si_2)$ for genus-2
surface, then those topological orders can be distinguished.  This supports the
conjecture proposed in \Ref{W9039}, that the representations of the mapping
class groups $\text{MCG}(\Si_g)$ for all genus-$g$ surfaces, plus the chiral
central charge, can fully characterize 2d topological orders.

\section{Probe and measure topological orders -- path integral approach}
\label{PMtopo}

In this section, we only consider bosonic systems.  Bosonic topological orders
can also be realized by path integral on triangulated spacetime $M^D$.  In this
section, we will discuss how to characterize topological orders using path
integral approach and the resulting partition functions.

\subsection{Topological partition function }

If the path integral is described by a well defined quantum field theory (such
as those that can be regularized by a tensor network path integral) taht has no
long range correlations, it will describe an anomaly-free
topological order.  But how to determine which anomaly-free topological order that
the path integral produces?  How to determine whether two path integrals give
rise to the same anomaly-free topological order or not?

One universal way to characterize all the topological orders is via the
partition function of the system.  In general, a  partition function on a
closed spacetime $M^D$ may have a form
\begin{align}
 Z(M^D) = \ee^{ -c_D L^D -c_{D-1} L^{D-1} -\cdots - c_0 L^0 - c_{-1}L^{-1}
-\cdots},
\end{align}
where $L$ is the linear size of $M^D$.  If the ground state does not contain
point-like, string-like, \etc defects, then $c_1=c_2=\cdots=c_{d-1}=0$.  In
this case, we can define a  volume-independent partition function via
\begin{align}
 Z^\text{top}(M^D) \equiv \lim_{L\to\infty} \frac{Z(M^D)}{\ee^{ -c_D L^D}} = \ee^{ - c_0 L^0}
\end{align}
When the calculated volume-independent partition function vanishes, it does not
mean the partition function vanishes. It just means that $c_i>0$, for some
$0<i<D$. This implies that the given space-time topology $M^D$ must contain
point-like, string-like, \etc topological excitations.  

We like to remark that it is not yet proven that using the above procedure to
define volume-independent partition function $Z^\text{top}(M^D)$ always works.
The assumption that $Z^\text{top}(M^D)$ can be well defined  equivalent to the
assumption that the partition function of a topological quantum field theory
can be defined via a microscopic path integral calculation.  However, in
\Ref{WW180109938}, it is shown that only some ratios of $Z^\text{top}(M^D)$ are
well defined.  For example we may choose triangulated manifold which contain
four disjoint pieces: $\t M = M_U\sqcup M_D \sqcup N_U \sqcup N_D$,  where the
boundaries of $M_U$, $ M_D$, $N_U$, and $N_D$ are simplicial complexes all
isomorphic to $B$.  Gluing the boundaries in different ways gives rise to $ M_1
=\bmm \includegraphics[height=0.45in]{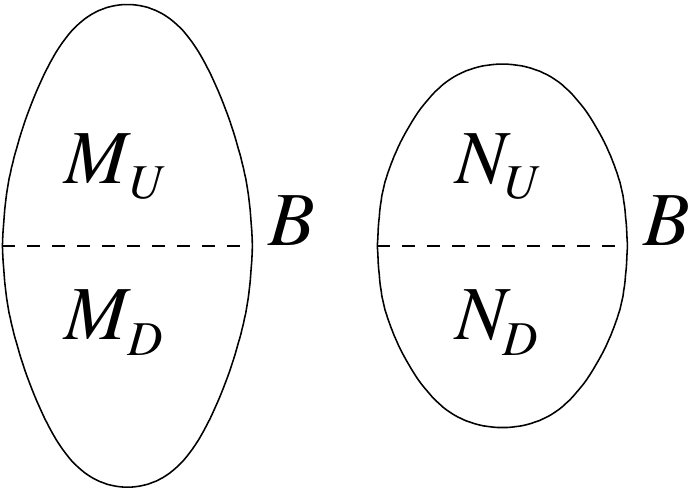} \emm $ and $ M_2 =\bmm
\includegraphics[height=0.45in]{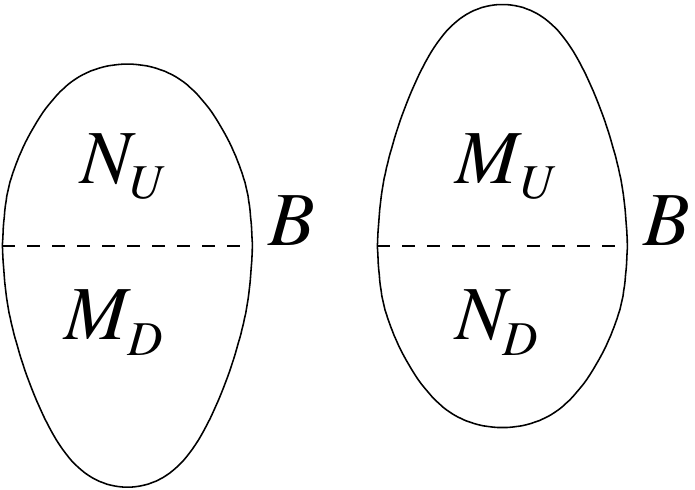} \emm$.  The  ratio of
$Z^\text{top}(M^D)$'s is given by
\begin{align}
&\ \ \ \
Z^\text{top}\Big( \bmm \includegraphics[height=0.5in]{MNB} \emm\Big)
\left/
Z^\text{top}\Big( \bmm \includegraphics[height=0.5in]{MNB1} \emm\Big) \right.
\nonumber\\
&=Z\Big( \bmm \includegraphics[height=0.5in]{MNB} \emm\Big)
\left/
Z\Big( \bmm \includegraphics[height=0.5in]{MNB1} \emm\Big) \right.
\nonumber\\
&=
\frac{
 Z(M_U \cup M_D) Z(N_U \cup N_D) 
}{
 Z(N_U \cup M_D) Z(M_U \cup N_D) 
}
\end{align}
since the volume term cancel exactly.  Such ratios are topological invariants.

\subsection{The gravitational Chern-Simons term and winding numbers}

To use partition function to characterize topological order, we consider
volume-independent partition function $Z^\text{top}(M^{D})$ on closed
space-time manifolds $M^{D}$.  To understand the universal structures in the
partition functions, let us use $\cM_{M^{D}}$ to denote the moduli space of the
closed space-time $M^{D}$ with different metrics but the same topology.  Then
the partition function $Z^\text{top}(-)$ can be viewed as a map from
$\cM_{M^{D}}$ to $\C$.
\begin{prop} 
If a path integral describes a short-range correlated systems, then its
volume-independent partition function on the moduli space $\cM_{M^{D}}$ of a
closed space-time $M^{D}$ is either always non-zero: $Z^\text{top}(M^{D}) \neq
0$, or always zero: $Z^\text{top}(M^{D}) = 0$. 
\end{prop}\noindent
\begin{proof}
If a volume-independent partition function is zero at some isolated points (or
regions) of the moduli space $\cM_{M^{D}}$, then a small local perturbation
will make it non-zero.  This will represent a diverging response, which should
not occur for  short-range correlated systems. Therefore,
$Z^\text{top}(M^{D})$ is either always non-zero or always zero.
\end{proof}
\noindent
So the non-zero volume-independent partition function $Z^\text{top}(\cdot)$ is
actually a map $Z^\text{top}: \cM_{M^{D}} \to \C -\{0\}\sim U(1)$.  If
$\pi_1(\cM_{M^{D}})\neq 0$, such map may have a non-trivial winding number.

Since $\pi_1(\cM_{M^{D}})=\text{MCG}(M^{D})$, the winding number is a group
homomorphism $\text{MCG}(M^{D}) \to \Z=\pi_1(U(1))$.  So the  winding numbers
(\ie the group homomorphisms) always form integer classes $\Z$. This leads us
to believe that the  winding numbers (or the group homomorphism
$\text{MCG}(M^{D}) \to \Z$) are always realized by the partition function
$Z^\text{top}(M^{D})$ that contains the gravitational Chern-Simons term
$\om_{D}$ 
\begin{align}
\label{Zom1}
 Z^\text{top}(M^{D})\sim \ee^{\ii \int_{M^{D}} \om_{D} }
\end{align}
where the gravitation Chern-Simnons term $\om_D$ is given by
\begin{align}
\label{omDP}
\dd \om_{D}=\sum \ka_{n_1n_2\cdots} P_{n_1n_2\cdots} ,
\end{align}
and $P_{n_1n_2\cdots}$ is a combination of Pontryagin classes which are the
only integer characteristic classes of oriented manifolds.  
Noticing that the winding numbers are given by
$\frac 1{2\pi} \int_{M^{D}\gext_f S^1} \dd \om_D $,
we have
\begin{prop} 
\begin{align}
\label{kagCS1}
 \int_{M^{D}\gext_f S^1} \frac{\dd \om_D}{2\pi}  =
 \sum \frac{\ka_{n_1n_2\cdots}}{2\pi} \int_{M^{D}\gext_f S^1} P_{n_1n_2\cdots}  \in \Z
\end{align}
for any mapping torus $M^{D}\gext_f S^1$ where
$Z^\text{top}(M^{D})$ is non zero.  
\end{prop} 
\noindent

Such type of winding numbers and the  partition function exist only when the
spacetime dimension $D=4k+3$.  We also note that there is always one
combination of Pontryagin classes for each $D=4k+3$  (corresponding to the
signature $\si$ of the manifold), whose value on mapping torus is always zero
(see Corollary 1.0.6. in \Ref{E0919}).  For such Pontryagin class, the
corresponding gravitational Chern-Simons term $\om_{D}^\si$ can have an
unquantized coefficient.  For example, in 2+1D, the naive consideration of
diffeomorphism invariance appear to require the gravitational Chern-Simons term
$\om_3$ to have a  coefficient that is quantized as $\ka=\frac{2\pi c}{24}$,
$c=0$ mod 8.  However, the above more careful consideration indicates that $c$
does not have to be quantized as $c=0$ mod 8.  This is consistent with the well
known fact that there are many 2+1D anomaly-free topological orders with $c$
different from $0$ mod 8, despite $c$ must
satisfy certain conditions (see Result \ref{cDgB} and \ref{cDgF}).

Clearly, two bosonic systems that give rise to partition functions with
different winding numbers must belong to two different phases.  So the  winding
numbers of partition functions are a type of topological invariants that can be
used to probe and measure the anomaly-free  topological orders.

\subsection{Beyond winding numbers}

To have an example of  topological orders with non-zero winding numbers,
we note that invertible topological order are described by volume-independent
partition functions that are pure $U(1)$ phase. 
\begin{prop} 
Since it has only trivial excitations, the volume-independent partition
function of an invertible topological order is non-zero for any closed
spacetime manifold $M^D$.
\end{prop} 
\noindent
In particular the $\Z$-class of invertible topological order 
(such as the $E_8$ quantum Hall state in $D=3$) are described by
\begin{align}
 Z^\text{top}(M^{D}) = \ee^{\ii \int_{M^{D}} \om_{D} }.
\end{align}
We have shown that $\ka_{n_1n_2\cdots}$ is quantized (see \eqn{omDP}) if
$P_{n_1n_2\cdots}$ is non zero on some mapping torus.

However, for invertible topological orders, $\ka_{n_1n_2\cdots}$ can be
quantized even if $P_{n_1n_2\cdots}$ is zero on any mapping torus.  To see
this, we need to consider more general ``loop'', \ie more general ``mapping
torus'', where the topology of the fiber $M^d$ can change as we go along the
loop.  In this case, the more general ``mapping torus'' can be any closed
$D+1$-dimensional manifold $M^{D+1}$.  Thus we require that
\begin{prop} 
for invertible bosonic topological orders
\begin{align}
 \ee^{\ii \int_{M^{D+1}} \dd \om_D } =
 \ee^{\ii \sum \ka_{n_1n_2\cdots} \int_{M^{D+1}} P_{n_1n_2\cdots} } =1
\end{align}
for any closed $M^{D+1}$,
which leads to a quantization of $\ka_{n_1n_2\cdots}$.
\end{prop} 
\noindent
We note that even $\om_{D}^\si$ is required to have a quantize coefficient.
For example, for 2+1D invertible topological orders
\begin{align}
\label{Zom2}
 Z^\text{top}(M^{3}) &= 
\ee^{\ii \ka \int_{M^{3}} \om_{3} }
=\ee^{\ii \frac{2\pi c}{24} \int_{M^{3}} \om_{3} }
\nonumber\\
&=\ee^{\ii \frac{2\pi c}{24} \int_{M^{4}} p_1 },\ \ \ \
\prt M^4=M^3,
\end{align}
where $c\equiv  12\ka/\pi$ must be quantized as 0 mod 8, since
$\int_{N^{4}} p_1 = 0$ mod 3 for closed 4-manifold $N^4$.  In fact $c$ is the
chiral central charge of the edge states and the above partition function
describes the stacking of $c/8$  $E_8$ quantum Hall states.

\subsection{Probe and measure the anomaly-free  topological orders
}  \label{PMclosedLtop}

In the last section, we have discussed the quantization of $\ka_{n_1n_2\cdots}$
for invertible topological orders.  For non-invertible topological orders, we
also have gravitational Chern-Simons terms, and have non-zero
$\ka_{n_1n_2\cdots}$.  In this case, $\ka_{n_1n_2\cdots}$ are also quantized.
However, the quantization condition are weaker, since the volume-independent
partition function is non-zero only on some sub-class of closed manifolds.  For
example, for bosonic topological orders with emergent fermions, the
volume-independent partition function vanishes on spacetime that is not spin.
The volume-independent partition function can be non-zero only on
spin-manifolds.  Only the spin-manifolds can impose the the quantization
conditions on $\ka_{n_1n_2\cdots}$.

Also for non-invertible topological orders, the volume-independent partition
function is not just a phase factor.
\begin{prop}
The non-invertible topological orders are characterized by the following topological invariants:\\
(1) Quantized  gravitational Chern-Simons terms (\ie quantized
$\ka_{n_1n_2\cdots}$'s) \\
(2) The absolution values of volume-independent partition function
$|Z^\text{top}(M^D)|$, on spacetime $M^D$ with vanishing Eular and Pontryagin
numbers (\ie $|Z^\text{top}(M^D)|$ is a constant on moduli space $\cM_{M^D}$).
\end{prop}

\subsection{How to probe and measure the boundary-gappable  topological orders
}  \label{PMexactLtop}

We know that an boundary-gappable  topological order can be described by a
topological path integral that is independent of retriangulation of space-time
and independent of local change of space-time metrics.  The topological path
integral directly give rise to the volume-independent partition function
$Z^\text{top}(M^{D})$, which is constant on $\cM_{M^{D}}$ locally.  Such a
topological path integral is a fixed-point of the renormalization group
transformation.  We propose that
\begin{prop} 
For boundary-gappable topological orders, we can use the volume-independent partition
function $Z^\text{top}(M^{D})$ of topological path integral to probe and
measure them.
\end{prop}\noindent
This conjecture has lead to some related researches and is confirmed for simple
boundary-gappable  topological orders.\cite{HW1339,MW1418,HMW1457} Since the topological
path integral is re-triangulation invariant, we see that $Z^\text{top}(M^{D})$
is not only independent of volume, it is also independent of shape. It only
depends on the topology of $M^{D}$.  Therefore, the volume-independent
partition function $Z^\text{top}(M^{D})$ is a topological invariant for
$D$-manifold $M^{D}$.  It might be even true that different boundary-gappable topological
orders give different topological invariants for at least some $M^{D}$'s.  In
2+1D,  the topological invariants from boundary-gappable  topological orders are the
Turaev-Viro invariants for 3-manifolds.\cite{TV9265}

We like to remark that the relation between volume-independent partition
functions $Z_\text{top}(M^{D})$ and boundary-gappable  topological orders is not
one-to-one.  Two volume-independent partition functions
$Z_\text{top}(M^{D})$ differ by a factor $W^{\chi(M^D)} \ee^{\ii \sum_{\{n_i\}}
\phi_{n_1n_2\cdots} \int_{M^D} P_{n_1n_2\cdots}}$ actually describe the same
topological order.  This is because the factor $W^{\chi(M^D)} \ee^{\ii
\sum_{\{n_i\}} \phi_{n_1n_2\cdots} \int_{M^D} P_{n_1n_2\cdots}}$ can be
produced by local counter terms, which are deformations within the same phase.

\subsection{Applications}

For bosonic 2+1D invertible topological orders, its volume-independent
partitions is non-zero in any closed orientable manifold $M^3$.  From the
quantization $\int_M^4 p_1 = 0$ mod 3, we find that the chiral central charge
$c=0$ mod 8.  For 2+1D non-invertible topological orders with emergent
fermions, its volume-independent partitions must be zero on any closed
orientable manifold $M^3$ that is not spin.  From the quantization $\int_{M^4}
p_1 = 0$ mod 48 for any spin manifold $M^4$, we see that $c$ only need to
satisfy $2 c n = 0$ mod 1 for a certain set of integers $n$.  Since the
partition function may not non-zero for all spin manifold, we cannot conclude
$2 c n = 0$ mod 1 for all integers.

The simplest class of 2+1D bosonic topological orders with emergent fermion has
quantum dimensions $(d_i) = (1,1,\sqrt2)$.  Their chiral central charge is
given by $c=\frac{2k+1}{2}$, $k\in \Z$.  We believe that the partition
functions for those topological orders are non-zero for any 2+1D spin manifolds
since those topological orders can be obtained by gauging the
fermion-number-parity in $p+\ii p$ superconductors in 2+1D.  So the chiral
central charge for those bosonic topological orders must satisfy $2 c n = 0$
mod 1 for all integers $n$, \ie $2 c = 0$ mod 1.  This is consistent with
$c=\frac{2k+1}{2}$.

There is a 2+1D topological order whose edge states are described $SU(2)$ level
6 conformal field theory with chiral central charge $c=\frac 94$.  Such a
topological order also has an emergent fermion.  Since the chiral central
charge does not satisfy $2 c = 0$ mod 1, the volume-independent partition
functions for such a topological order cannot be non-zero on all spin
manifolds.  It is an open problem to understand on which class of manifolds
that the volume-independent partition functions are non-zero.

~

\noindent {\bf Acknowledgement}: LK is supported by the Science, Technology and Innovation Commission of
Shenzhen Municipality (Grant No. ZDSYS20170303165926217) and Guangdong
Provincial Key Laboratory (Grant No.2019B121203002) and NSFC under Grant No.
11971219.  XGW is partially supported by NSF DMS-1664412 and by the Simons
Collaboration on Ultra-Quantum Matter, which is a grant from the Simons
Foundation (651440).

\bibliography{../../bib/wencross,../../bib/all,../../bib/publst} 

\end{document}